\documentclass[sigconf]{acmart}
\usepackage{graphicx}
\usepackage{amsmath}
\usepackage{amsthm}
\usepackage{booktabs}
\usepackage{algorithmic}
\usepackage{subfigure}
\usepackage{multirow}
\usepackage{color}
\usepackage{bm}

\theoremstyle{definition}

\usepackage{multirow}
\usepackage{amsopn}
\usepackage{makecell}
\usepackage{booktabs}
\usepackage{amsfonts}
\usepackage[ruled,linesnumbered]{algorithm2e}
\usepackage{enumitem}
\usepackage{balance}
\usepackage{diagbox}

\renewcommand\footnotetextcopyrightpermission[1]{} 
\AtBeginDocument{%
  \providecommand\BibTeX{{%
    \normalfont B\kern-0.5em{\scshape i\kern-0.25em b}\kern-0.8em\TeX}}}

\begin{document}

\title{Multi-Task Fusion via Reinforcement Learning for Long-Term User Satisfaction in Recommender Systems}

\author{Qihua Zhang}
\affiliation{
  \institution{Tencent Inc.}
  \city{Shenzhen}
  \country{China}
  }
\email{kirahzhang@tencent.com}

\author{Junning Liu}
\affiliation{
  \institution{Tencent Inc.}
  \city{Beijing}
  \country{China}
  }
\email{korchinliu@tencent.com}

\author{Yuzhuo Dai}
\affiliation{
  \institution{Tencent Inc.}
  \city{Shenzhen}
  \country{China}
  }
\email{zoyadai@tencent.com}

\author{Yiyan Qi}
\affiliation{
  \institution{Tencent Inc.}
  \city{Shenzhen}
  \country{China}
  }
\email{yiyanqi@tencent.com}

\author{Yifan Yuan}
\affiliation{
  \institution{Tencent Inc.}
  \city{Shenzhen}
  \country{China}
  }
\email{yaphetyuan@tencent.com}

\author{Kunlun Zheng}
\affiliation{
  \institution{Tencent Inc.}
  \city{Beijing}
  \country{China}
  }
\email{kunlunzheng@tencent.com}

\author{Fan Huang}\authornote{Corresponding author}
\affiliation{
  \institution{Tencent Inc.}
  \city{Shenzhen}
  \country{China}
  }
\email{sinohuang@tencent.com}

\author{Xianfeng Tan}
\affiliation{
  \institution{Tencent Inc.}
  \city{Beijing}
  \country{China}
  }
\email{victan@tencent.com}

\begin{abstract}
Recommender System (RS) is an important online application that affects billions of users every day. The mainstream RS ranking framework is composed of two parts: a Multi-Task Learning model (MTL) that predicts various user feedback, i.e., clicks, likes, sharings, and a Multi-Task Fusion model (MTF) that combines the multi-task outputs into one final ranking score with respect to user satisfaction. There has not been much research on the fusion model while it has great impact on the final recommendation as the last crucial process of the ranking. To optimize long-term user satisfaction rather than obtain instant returns greedily, we formulate MTF task as Markov Decision Process (MDP) within a recommendation session and propose a Batch Reinforcement Learning (RL) based Multi-Task Fusion framework (BatchRL-MTF) that includes a Batch RL framework and an online exploration. The former exploits Batch RL to learn an optimal recommendation policy from the fixed batch data offline for long-term user satisfaction, while the latter explores potential high-value actions online to break through the local optimal dilemma. With a comprehensive investigation on user behaviors, we model the user satisfaction reward with subtle heuristics from two aspects of user stickiness and user activeness. Finally, we conduct extensive experiments on a billion-sample level real-world dataset to show the effectiveness of our model. We propose a conservative offline policy estimator (Conservative-OPEstimator) to test our model offline. Furthermore, we take online experiments in a real recommendation environment to compare performance of different models. As one of few Batch RL researches applied in MTF task successfully, our model has also been deployed on a large-scale industrial short video platform, serving hundreds of millions of users.
\end{abstract}




\maketitle

\section{Introduction}
With the information explosion on the Internet, Recommender Systems (RS) that aim to recommend potentially interesting items for users, are playing an increasing important role in various platforms including E-commerce sites \cite{linden2003amazon,zhou2018deep,zou2019reinforcement,gu2020hierarchical}, videos sharing sites \cite{covington2016deep}, social networks \cite{he2014practical,gu2016hlgps}, etc. 
There are usually two main stages in industrial RS: candidate generation and ranking \cite{zhao2019recommending}.
The first stage selects hundreds or thousands of candidates from millions or even billions of items,
while the second stage returns several top-ranked items for each user from the candidates.

Given a user query, the quality of the ranking results is a leading factor in affecting user satisfaction. 
Early works \cite{zhou2018deep,zhou2019deep,ouyang2019deep,pi2019practice} usually only consider a single instant metric, e.g., Click-Through Rate (CTR), and train a ranking model over this metric.
In practice, it is usually hard to measure real user satisfaction with just one metric.
For example, in news recommendation scenario, users may click the recommended news with click bait like the cover pictures or 
titles but quickly exit when they are not interested in the content.
Clearly, only considering CTR will cause a lot of improper recommendations and this encourages us to explore multiple metrics through different user behaviors.
When more than one metric is considered, an essential question arises, as to how should these metrics be combined to optimize ranking quality.

Model fusion is a popular approach to solve this problem \cite{dong2010towards,carmel2020multi}, which is usually composed of two parts: 
1) a Multi-Task Learning model (MTL) \cite{ma2018modeling,tang2020progressive} that predicts multiple metrics associated with user satisfaction;
2) a Multi-Task Fusion model (MTF) \cite{pei2019value,han2019optimizing} that constructs a combination function based on those predicted scores and produces the final ranking.
Compared with the MTL works on the first step, MTF algorithms are crucial for recommendation quality but there has been little good research on it.

In this paper, we focus on the MTF algorithms in RS. Given a fusion function $f(o_1,o_2,\ldots,o_k)$ with respect to ranking scores $o_1,o_2,\ldots,o_k$ from different predicted tasks, a naive way to find out the optimal fusion weights is to use hyper-parameter searching techniques such as Grid Search and Bayesian Optimization \cite{movckus1975bayesian}.
These methods fail in large RS not only because of their inefficiency but also because they can't produce personalized weights for different users and different contexts to make accurate recommendations.
One can solve these issues by bridging user states and fusion weights through neural networks and turn to optimize the network weight via Evolutionary Strategy \cite{beyer2002evolution}.
However, all the above methods still focus on optimizing instant user satisfaction in a greedy way but ignore long-term user satisfaction. To reduce the expected regret of recommendations and improve the long-term rewards for RSs, we need to consider instant user satisfaction as well as delayed user satisfaction brought by the long-term utility of the recommendation. After all, the current recommendation may affect user preference later.

Intuitively, Reinforcement Learning (RL) is usually designed to maximize long-term rewards.
However, applying RL in the large-scale online RS to optimize long-term user satisfaction is still a non-trivial problem. 
1) The long-term user satisfaction is very complicated and can be measured in various user behaviors.
How to build feasible reward according to different behaviors is challenging.
2) In order to learn an optimal RL policy effectively, the recommender agent requires a very large number of sequential interactions with real users to trial and error. However, on-policy RL would harm user experiences with the noise generated by online exploration during learning.
3) An alternative is to build a recommender agent offline through the logged data, which can mitigate the cost of the trial-and-error exploration. 
Unfortunately, traditional off-policy methods not only suffer from the Deadly Triad problem \cite{thrun2000reinforcement}, i.e., the problem of instability and divergence arises whenever combining function approximation, bootstrapping, and offline training, but also suffer from serious extrapolation error\cite{fujimoto2019off} where state-action pairs not in the fixed batch data, also called out-of-distribution (OOD) training data, are erroneously estimated to have unrealistic values.

Considering the above problems, we propose a Batch RL based Multi-Task Fusion framework (BatchRL-MTF) to optimize long-term user satisfaction. Our model consists of two components: 1) Batch RL framework learns an optimal recommendation policy with high returns, strong robustness and less extrapolation error offline, which provides the fusion function with a set of personalized fusion weights trading off instant and long-term user satisfaction; 2) Online Exploration Policy interacts with real users online to discover high-reward fusion weights as offline batch samples. 
In particular, we formulate our MTF task as Markov Decision Process (MDP) within a recommendation session to model the sequential interaction between the user and the RS.
We first comprehensively investigate different user behaviors, and deliberately design the reward based on these behaviors to optimize long-term user satisfaction from two aspects of user stickiness and user activeness.
To reduce the learning costs of our model and the damage to user experience, we present Batch RL to enhance learning efficiency, mitigate extrapolation error and optimize accumulated rewards according to the history logs.
In addition, we conduct online exploration that can find more high-value state-action pairs to ensure the sufficient training of our model, which not only reduces extrapolation error but also prevents our model from falling into local optimum.
In experiments, we propose a conservative offline policy estimator (Conservative-OPEstimator) to verify the outstanding performance of the model by evaluating the long-term rewards brought by it. 
Meanwhile, we conduct online evaluations for BatchRL-MTF with competitive baselines on a real-world short video platform. 
The significant improvements on user stickiness and user activeness exhibit the effectiveness of our BatchRL-MTF in practice.
The major contributions of our work include:

\begin{itemize}[leftmargin=*]
\item  We formulate the session-based MTF task as an MDP and exploit Batch RL to optimize long-term user satisfaction, which is one of the few works of Batch RL applied in MTF task successfully.

\item We design our reward function with respect to multiple user behaviors that relate to user stickiness and user activeness, and train our BatchRL-MTF based on history logs. 
Specially, our BatchRL-MTF contains two components: Batch RL framework and online exploration policy. 
The former learns an optimal recommendation policy offline and the latter explores potential high-value state-action pairs online.
We show that our framework can mitigate the deadly triad problem and extrapolation error problem of traditional off-policy applied in practice RSs.

\item We creatively propose a conservative offline policy estimator (Conservative-OPEstimator) to test our model offline, while conducting online experiments in real recommendation environment to demonstrate our model outperforms baselines greatly. In addition, BatchRL-MTF has been deployed in the short video recommendation platform and remarkably improved 2.550\% app dwell time and 9.651\% user positive-interaction rate.
\end{itemize}
\section{Problem Formulation}
\begin{figure}[t]
    \centering
    \includegraphics[width=8cm,height=5cm]{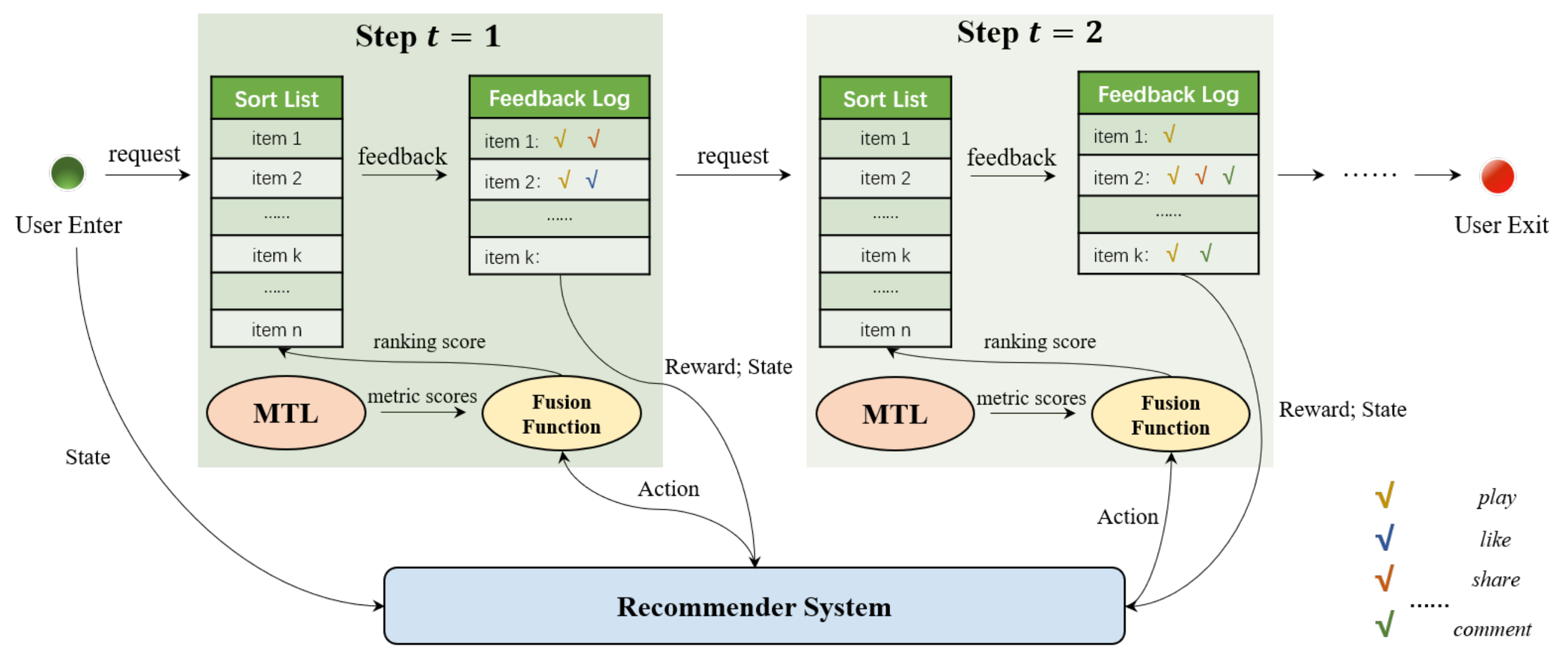}
    \captionsetup{font={small,bf},justification=raggedright}
    \caption{The User-RS interaction within a session-based recommendation.}
    \label{fig:session}
\end{figure}
As shown in Figure \ref{fig:session}, in the short video recommendation scenario, the RS agent interacts with a user $u \in U$ at discrete time steps within a recommendation session.
At each time step $t$, the RS feeds a top-ranked item $i^{(t)} \in I$ (or item list $(i_1^{(t)},\dots,i_l^{(t)})$) to user $u$ and receives feedback vector ${v}^{(t)} = (v_1^{(t)},\ldots,v_m^{(t)})$, where $I$ is the candidate set and $v_i^{(t)}$, $1 \le i \le m$ is a specific user's behavior (e.g., video play time, like, sharing, etc.) on item $i^{(t)}$.
The mainstream RS ranking pipeline is composed of two parts: an MTL model that predicts various user behaviors and an MTF model that combines the multi-task outputs, i.e., ${o} = (o_1,\ldots,o_k)$, into one final ranking score. We employ Progressive Layered Extraction (PLE) model \cite{tang2020progressive} with excellent performance to conduct multi-task predictions.
To output the personalized recommendation matching user preferences, we build a fusion function $f({o})$ that aggregates these predicted scores to model user satisfaction.
Considering the effect of magnitude difference among $o_1,\ldots,o_k$, we define the function as follow:
\begin{equation}\label{eq:proxyfunction}
    f({o} | {\alpha})= \sum\limits_i^k \alpha_i \log(o_i+\beta_i),
\end{equation}
where ${\alpha}=(\alpha_1, \ldots, \alpha_k)$ is the fusion weights to be optimized and ${\beta}=(\beta_1, \ldots, \beta_k)$ is a constant bias set by priori knowledge to smooth those $o_1,\ldots,o_k$. 
In this paper, we aim to find out the optimal weights ${\alpha}$ that maximize long-term user satisfaction. 

In particular, we study the above problem as Markov Decision Process (MDP) within the recommendation session, where an agent (the RS) interacts with the environment (users) by sequentially recommending items over time, to maximize the cumulative reward within the session. 
Formally, the MDP consists of a tuple of five elements $(\mathcal{S},\mathcal{A},\mathcal{P},\mathcal{R},\gamma)$:

$\bullet$ \textbf{State space $\mathcal{S}$}: is a set of user state $s$, which is composed of two parts: user profile feature (e.g., age, male, location, etc.) and user's interaction history feature (e.g., like, sharing, skipping, etc.). In our RS, the latter is usually formulated by a user's various interactions with his/her watched last 500 videos. 

$\bullet$ \textbf{Action space $\mathcal{A}$}: The action determines the ranking of the candidate list.
In our problem, the action $a$ is the fusion weight vector ${\alpha}= (\alpha_1,\dots,\alpha_k)$ to be optimized in Equation (\ref{eq:proxyfunction}). With the fusion weights, the agent can further calculate the final ranking scores and return recommendation item(s).

$\bullet$ \textbf{Reward $\mathcal{R}$}: After the RS takes an action $a_t$ at the state $s_t$, i.e., figure out the fusion scores and return a recommendation item to response user's request at time step $t$, a user will feedback with different behaviors.
We define the instant reward $r(s_t, a_t)$ based on these behaviors.

$\bullet$ \textbf{Transition probability $\mathcal{P}$}: Transition probability $p(s_{t+1} | s_t, a_t)$ defines the state transition from $s_{t}$ to $s_{t+1}$ after taking action $a_t$. 
As the user state is defined as user profile and his/her interaction history, the next state $s_{t+1}$ is determined after user feedback.

$\bullet$ \textbf{Discount factor $\gamma$}: Discount factor $\gamma \in [0,1]$ is introduced to control the present value of future reward.

With the above notations and definitions, the task of Batch RL applied in MTF can be defined as: given the interaction history logs between RS agent and users in MDP form within the recommendation session, how to find a policy obtaining the optimal fusion weights ${\alpha}$ to maximize the cumulative reward of RS.
In the rest of this paper, we omit the superscript $(t)$ for simplicity.

\section{Proposed Framework}\label{sec:method}

In this section, we propose a Batch RL framework for short video RSs to optimize long-term user satisfaction. 
Specifically, we design the reward based on multiple user behaviors and present our Batch RL architecture. Firstly, we discuss how to train our BatchRL-MTF offline via users' behavior log.
Secondly, we also conduct online exploration to discover potential high-value state-action pairs in real recommendation environment.
More importantly, we briefly describe the implementation of the proposed model on a real-world short-video recommender platform.

\subsection{Reward Function}
In a recommendation session, the RS interacts with users in turn, i.e., the agent takes an action $a$ at the state $s$ (i.e., recommending a video to a user with respect to MTL ranking scores) and then the user provides his/her feedback $v$.
To comprehensively measure the instant satisfaction, we formulate the reward function as follow:
\begin{equation}\label{eq:reward}
    r(s,a)= \sum\limits_{i=1}^{m} w_{i} v_{i},
\end{equation}
where $w_i$ is the weight of feedback $v_i$.
In our RS, feedback $v_1, \ldots, v_m$ includes video play time, play integrity and interaction behavior such as liking, sharing, commenting, liking, etc.
And weights $w_1, \ldots, w_m$ are set via extensive statistic analysis on relationships between these feedback and future user app dwell time.

\subsection{Batch Reinforcement Learning for MTF}
Due to costly real-world interactions, it is difficult to apply on-policy and off-policy algorithms in large-scale industrial RSs. 
An alternative approach is learning an RL-based recommender agent solely from historical logs, which is also known as Batch Reinforcement Learning. 
However, a severe problem in Batch RL approaches is extrapolation error, which is caused by the distribution difference between training data and learned policy.
To solve this problem, we exploit Batch-Constrained
deep Q-learning model (BCQ) \cite{fujimoto2019off}, based on the popular Actor-Critic architecture, to learn the optimal fusion weights in our framework.


\subsubsection{Actor Network.}\label{sec:actor}
The Actor network includes two sub-network: (a) the action generative network $G_{\theta}=\{E_{\theta_1},D_{\theta_2}\}$ and (b) the action perturbation network $P_{\omega}(s,a,\rho)$.

Specially, network $G_{\theta}=\{E_{\theta_1},D_{\theta_2}\}$ is a conditional variational auto-encoder (VAE) to generate candidate action set of which the distribution is similar to that of training samples $\mathcal{B}$, which reduces extrapolation error and enhances our framework robustness.
$G_{\theta}$ can be further divided into an encoder block $E_{\theta_1}(z|s,a)$ and a decoder block $D_{\theta_2}(a|s,z)$.
To be more specific, encoder $E_{\theta_1}(z|s,a)$ learns the latent distribution of $(s,a) \in \mathcal{B}$ and forces it to approximate $\mathcal{N}(0,1)$ via KL divergence.
$E_{\theta_1}(z|s,a)$ has two outputs, i.e., the mean value $\mu$ and standard deviation $\delta$.
Decoder $D_{\theta_2}(a|s,z)$ takes the latent vector $z \sim \mathcal{N}(\mu,\delta^2)$ and the user state $s$ as inputs to output the action $\hat{a}$ that is similar to $(s,a)\in \mathcal{B}$.
Formally, we train the VAE based on the following objective on the log-likelihood of the dataset:
\begin{equation}\label{eq:VAE}
    \theta\gets
    \mathop{argmin}_{\theta}\sum\limits_{(s,a) \in \mathcal{B}}(a - G_{\theta}(s))^2+\mathop{D_{KL}}_{(s,a) \in \mathcal{B}}(E_{\theta_1}(z|s,a)| \mathcal{N}(0,1)).
\end{equation}

\begin{figure}[t!]
    \centering
    \includegraphics[width=8cm,height=2cm]{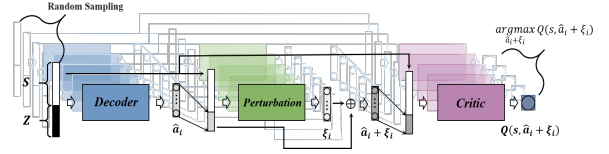}
    \captionsetup{font={small,bf},justification=raggedright}
    \caption{The BatchRL-MTF policy.}
    \label{fig:policy}
\end{figure}

To increase the diversity of the actions from VAE, we use network $P_{\omega}(s,a,\rho)$ to generate a perturbation $\xi \in [-\rho, \rho]$ with respect to state $s$ and action $a$.
Specifically, given user state $s$, we generate $n$ actions $\{{ \hat{a}_{i} \sim G_{\theta}(s)}\}_{i=1}^n$ as candidates. 
Meanwhile, $P_{\omega}(s,a,\rho)$ generates $n$ perturbations $\{\xi_{i}\}_{i=1}^n$, $\xi_{i} = P_{\omega}(s, \hat{a}_i ,\rho)$ that updates the actions as $\hat{a}_{i} + \xi_i$.
As shown in Figure \ref{fig:policy}, the optimal action is selected from $n$ perturbed actions as:
\begin{equation}
    \pi(s)=\mathop{argmax}_{\hat{a}_{i}+\xi_{i}} Q(s,\hat{a}_{i}+\xi_{i}),
\end{equation}
and $P_{\omega}(s,a,\rho)$ is optimized through the deterministic policy gradient algorithm \cite{silver2014deterministic}:
\begin{equation}\label{eq:Perb}
    \omega \gets \mathop{argmax}_{\omega}\sum\limits_{s \in \mathcal{B}\atop \hat{a} \sim G_{\theta}(s)} Q(s, \hat{a}+P_{\omega}(s, \hat{a} ,\rho)).
\end{equation}

\subsubsection{Critic Network.}
The Critic network $Q_{\phi}(s,a)$ aims to estimate the cumulative reward of a state-action pair $(s,a)$. 
Following the common setting, we build four Critic networks during the learning process, i.e., two current networks $Q_{\phi_1},Q_{\phi_2}$ and two target networks $Q_{\phi'_1},Q_{\phi'_2}$. 
Specifically, the goal of Critic network is to minimize TD-error in the bootstrapping way:
\begin{equation}\label{eq:Q}
    \phi_{j} \gets \mathop{argmin}_{\phi_{j}}  \sum\limits_{(s,a,s') \in \mathcal{B} } [y-Q_{\phi_{j}}(s,a)]^2, \quad j\in\{1,2\},
\end{equation}
and the learning target $y$ is set according to the Clipped Double Q-Learning \cite{fujimoto2018addressing} that reduces the overestimation bias:
\begin{equation}
\begin{split}
    {y}= &\;r+\gamma\mathop{max}_{a'}[\mathop{min}_{j=1,2}Q_{\phi'_{j}}(s',a')],
    \\&a'\in \{\hat{a_{i}'}  + P_{\omega'}(s', \hat{a_{i}'} ,\rho), \hat{a_{i}'}  \sim G_{\theta}(s') \}_{i=1}^n
\end{split}
\end{equation}
where $a'$ sampled from the generative model and outputted by the target action perturbation network.

\begin{algorithm}[t]
	\SetKwInOut{Input}{Input}
	\SetKwInOut{Output}{Output}
	\BlankLine
 	\Input{the transition dataset $\mathcal{B}$, mini-batch size $M$, number of sampled actions $n$, perturbation bound $\rho$, target network update rate $\eta_{t}$, discount factor $\gamma$, delay update step $L$, number of training epochs $EP$}
 	\BlankLine
	
	Initialize generative network $G_{\theta}=\{E_{\theta_1},D_{\theta_2}\}$, perturbation network $P_{\omega}$, and Critic network $Q_{\phi}=\{Q_{\phi_{1}},Q_{\phi_{2}}\}$ with random parameters $\theta,\omega,\phi$\;
	
	Initialize target networks $P_{\omega'}$, $Q_{\phi'}=\{Q_{\phi_{1}'},Q_{\phi_{2}'}\}$ with $\omega'\gets\omega,\phi_{1}'\gets \phi_{1},\phi_{2}'\gets \phi_{2}$\;
	
	
	\ForEach{$1 \le ep \le Ep$}{
	    Sample mini-batch of $M$ transitions $(s,a,r,s')$ from $\mathcal{B}$ randomly\;
	    Update $G_{\theta}$ according to Equation (\ref{eq:VAE})\;
	    Sample $n$ actions
	    $\{\hat{{a_{i}'}}  \sim G_{\theta}(s')\}_{i=1}^n$\;
	    Generate $n$ perturbed actions $\{\hat{a_{i}'}  + P_{\omega}(s', \hat{a_{i}'}  ,\rho)\}_{i=1}^n$\;
	    Update $P_{\omega}$ according to Equation (\ref{eq:Perb})\;
	    Update $Q_{\phi}$ according to Equation (\ref{eq:Q})\;
	    \If{$ep \% L == 0$}{
	        $\omega'\gets\eta_t\omega+(1-\eta_t)\omega'$\;
	        $\phi_{i}'\gets\eta_t\phi_i+(1-\eta_t)\phi_i'$, $i=1,2$\;
	    }
	}

	\caption{Offline Training of BatchRL-MTF.\label{alg:A}}
\end{algorithm}

\subsubsection{Offline Model Training}
We train our Batch RL model offline based on the pre-collected dataset and the learning algorithm is shown in Algorithm \ref{alg:A}.
We first construct a transition dataset $\mathcal{B}$ based on history trajectories that record the online interaction between the user and the recommender agent. 
To improve the utilization efficiency of samples and speed up the convergence of our model, we used replay buffer to store historical transitions during training process. 
Specially, replay buffer is a fixed-size queue, where new transitions will replace old ones.
Based on the mini-batch sampled from $\mathcal{B}$, we update sub-networks $G_{\theta}$, $P_{\omega}$, and $Q_{\phi}$ in order.
In addition, we perform soft update on target networks every $L$-iteration to reduce  overestimation bias.

\subsection{Online Exploration}
Although the proposed Batch RL framework can reduce extrapolation error, we have considered the Actor network learns a fine-tuned policy based on training data generated by behavior policy. 
As a result, the agent’s performance is limited by behavior policy. If behavior policy doesn't explore the real environment enough, it will cause extrapolation error and local optimum problems in target policy.
To solve this problem, we propose an online exploration policy to discover potential high-reward state-action pairs in online serving.
In particular, we first perform two types of exploration on two groups of users separately.

$\bullet$ \textbf{Random Exploration.} The agent randomly samples an action from Gaussian distribution to interact with real users while having no priori knowledge.

$\bullet$ \textbf{Action-Noise Exploration.} 
To improve exploration efficiency and exploit the priori knowledge of optimal actions, we conduct action-noise exploration policy, which adds Gaussian noise to the action outputted by the optimal target policy.
Formally, we have:
\begin{equation}
    \pi_{ep}(s)=\pi_{t}^*(s)+\epsilon,\ \epsilon \sim \mathcal{N}(0,0.1).
\end{equation}
where $\pi_{ep}(s)$ is exploration policy and $\pi_{t}^*(s)$ is the target policy of current model.
The random exploration policy can enrich the diversity of actions, which prevents the local optimum problem; 
the action-noise exploration policy makes full use of the priori knowledge of optimal actions to explore nearby actions with potentially high value.
To exploit the advantages of these two exploration policy, we propose an online exploration policy that combines them, called \emph{Mixed Multi-Exploration policy}.

\begin{figure}[t!]
    \centering
    \includegraphics[width=8cm,height=2cm]{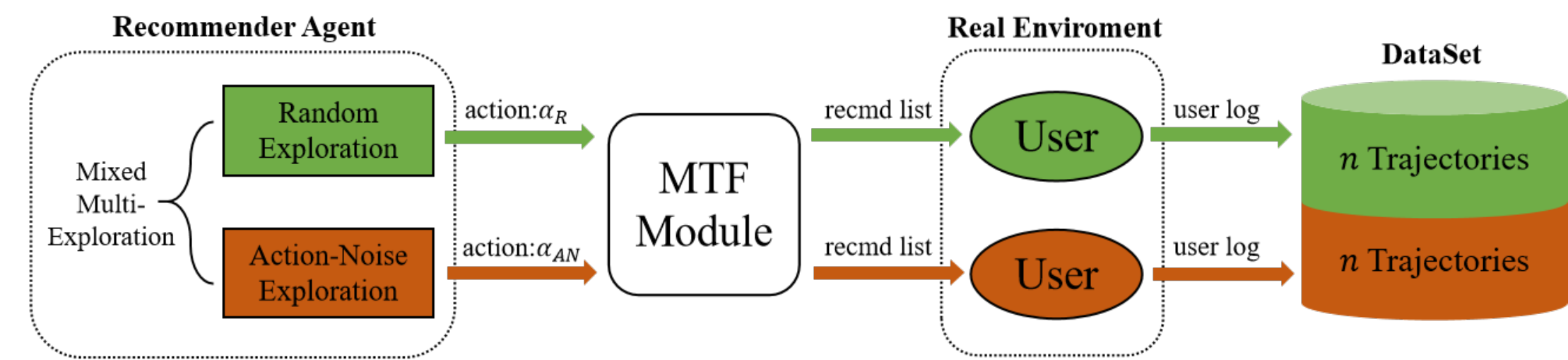}
    \captionsetup{font={small,bf},justification=raggedright}
    \caption{The online exploration policy of BatchRL-MTF framework.}
    \label{fig:online exploratioon}
\end{figure}

$\bullet$ \textbf{Mixed Multi-Exploration} is composed of random exploration and action-noise exploration, shown in Figure \ref{fig:online exploratioon}. It constructs the training dataset with equal amounts of trajectory samples collected by random exploration and action-noise exploration.

Different from exploration with simulators, online exploration will receive feedback from real users, which results in more accurate and unbiased rewards.
In experiments, we will show that training model with the dataset from our mixed multi-exploration policy will exhibit better performance.



\begin{figure}[t!]
    \centering
    \includegraphics[width=8cm, height=6cm]{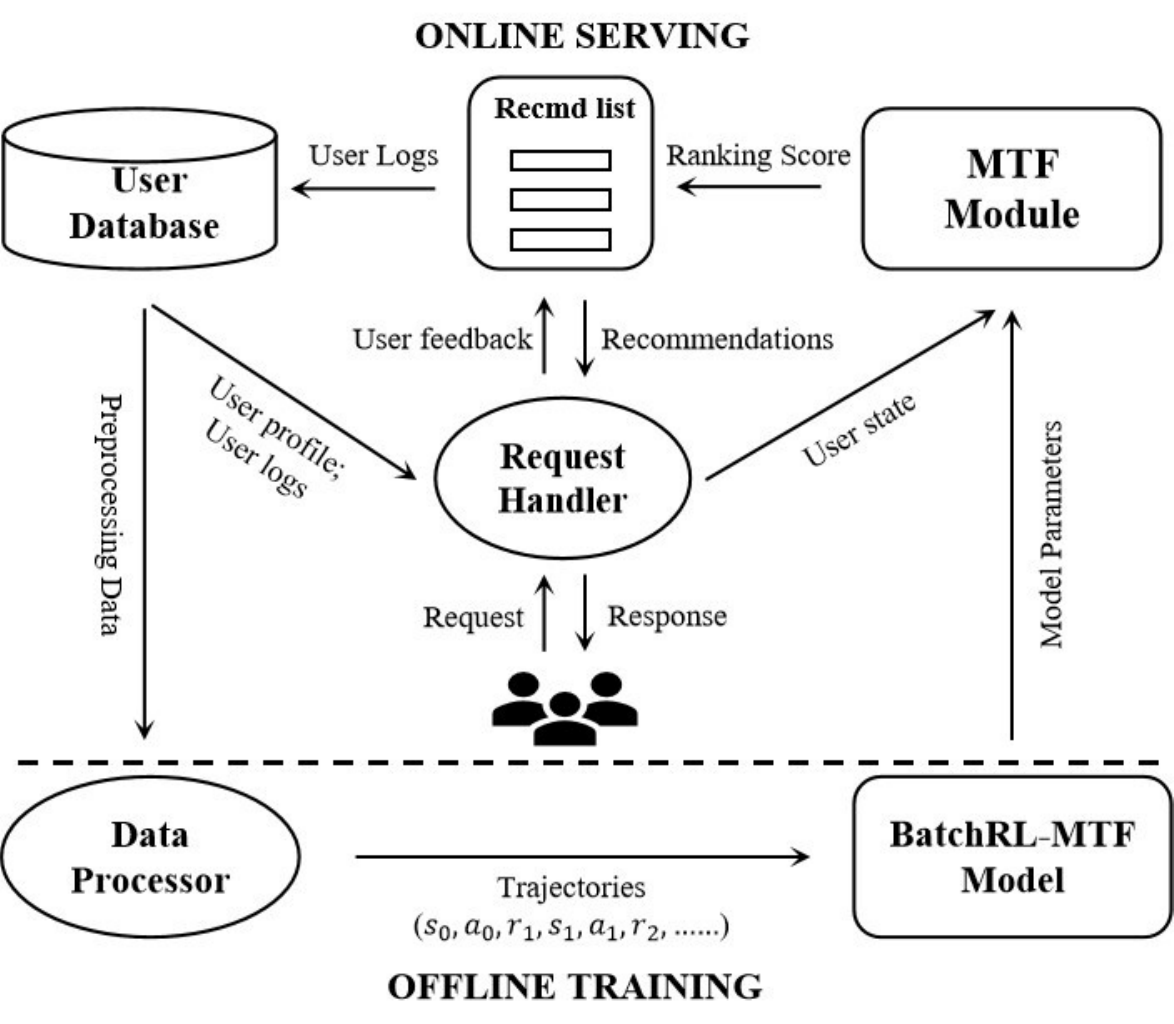}
    \captionsetup{font={small,bf},justification=raggedright}
    \caption{BatchRL-MTF framework of our short video recommender system.}
    \label{fig:galaxy}
\end{figure}
\subsection{Short Video Recommender System with BatchRL-MTF}
We implement our model on a large-scale short video recommendation platform.
As shown in Figure \ref{fig:galaxy}, the RS is composed of two components: offline training and online serving.
The two components are connected through user database and online MTF module, which are used for collecting user-agent interaction logs and conducting MTF task based on the BatchRL-MTF model.

The offline training component is responsible for data preprocessing and BatchRL-MTF model training. 
To be more specific, data processor pulls recent user data including user profile and user history logs from user database and organises these data as interaction trajectories $\{\tau_i = ( s^{(0)}_i, a^{(0)}_i, r^{(1)}_i,s^{(1)}_i,a^{(1)}_i,r^{(2)}_i,...)\}_{i=1}^N$, where each trajectory represents a user-agent interaction session. 
Then, we exploit Algorithm \ref{alg:A} to learn a new model based on these trajectories and update online model.
In practice, we re-train our model daily based on the trajectory data collected in the past three days. 

The online serving component provides personalized recommendations and collects user data. 
When receiving a user request, request handler retrieves the relevant information of this user from user database,
constructs user state feature, and transfers user state to MTF module. 
According to the user state, the MTF module produces a final ranking score for each video candidate.
Finally, request handler feeds top-ranked recommended video(s) to the user and collect his/her feedback.

\section{Experiments}\label{sec:experiment}

\subsection{Dataset}\label{subsec:Experiment dataset}
Our batch data is collected from a real-world short video recommendation platform, including about 3.142 million sessions and 11.155 million user-agent interactions.
For offline experiments, to ensure that models fully learn the sequential information between recommendation sessions, we take the first 90\% user sessions of the dataset in time order as the training dataset $\mathcal{B}$ to train the model and the remaining 10\% user sessions as testing dataset $\mathcal{D}$ to evaluate the model. For online experiments, we exploit the batch dataset to learn an optimal model and deploy it in our online short video recommendation platform for one month to conduct A/B tests.

\subsection{Implementation Details}\label{subsec:Experiment setup}
In BatchRL-MTF, the input user state $s$ is concatenated by user profiles feature and interaction feature from last $500$ watched videos. 
Its output action $a$ is $12$-dimensional vector representing fusion weights ${\alpha}$ in Equation (\ref{eq:proxyfunction}).
All networks in BatchRL-MTF are MLP with ReLU activation function in hidden layers and are optimized based on Adam optimizer.
For action perturbation network $P_{\omega}$, we use Tanh activation function maps its output $\psi$ to $[-1,1]$ and let the perturbation bound $\rho = 0.15$.
We set the reward discount factor as $\gamma = 0.95$.
The initial learning rate $\eta$ for the action generative network, the action perturbation network and critic networks is set to $0.1 \times 10^{-2}$, $0.1 \times 10^{-3}$ and $0.2 \times 10^{-3}$ separately; the soft update rate and the delay update step for target networks are $\eta_t = 0.05$ and $\emph{L} = 10$ separately. The above parameters are determined by offline experiments for maximizing long-term returns.
In addition, the replay buffer size, mini-batch size and training epochs in our training process are set to $100,000$, $M = 256$, and $Ep = 300,000$ respectively.


\subsection{Evaluation Setting}
\subsubsection{Offline policy estimator.}
Online A$/$B testing, a general method in the industry for evaluating new recommendation technologies, is also one of our preferred methods in comparison experiments. But it takes time and costs resources. Most importantly, the terrible policy could hurt user experiences. To overcome the above problems and accelerate iteration of new technologies, we propose an offline policy estimator to offline evaluate the performance of RL model. Inspired by Fitted Q Evaluation (FQE) algorithm \cite{le2019batch,voloshin2019empirical} and Conservative Q-Learning (CQL) algorithm \cite{kumar2020conservative}, our Conservative Offline Policy Estimator, also called Conservative-OPEstimator, is designed as:
\begin{equation}\label{eq:OPE Q value}
    \hat{V}(\pi_e) = \dfrac{1}{n}\sum_{i=1}^{n}\sum_{s^0_i \sim d^{0}(s)  \atop a \sim \pi_{e}(a|s) }\pi_{e}(a|s^0_i)\hat{Q}(s^0_i,a,\theta)
\end{equation}
where $d^{0}(s)$ is the initial state distribution; $\pi_{e}(a|s)$ is the policy to be evaluated; $\hat{Q}(\cdot,\theta)$ is used to estimate how many present values of long-term revenues within an online real recommendation session will be produced by an initial state-action pair. To improve the accuracy of estimation, we resort to CQL algorithm to construct $\hat{Q}(\cdot,\theta)$ by function approximation, which punishes the estimated Q values of state-action pairs not in the dataset $\mathcal{D}$ to prevent overestimation of the policy value. $\hat{Q}(\cdot,\theta)=\lim\limits_{k \rightarrow \infty}\hat{Q}_{k}$ where:
\begin{equation}\label{eq:OPE Q}
\begin{split}
    \hat{Q}_{k+1} &\gets  \mathop{argmax}_{\theta} \alpha \cdot \mathcal{R}(\theta) + \frac{1}{2} \cdot \mathbb{E}_{s,a,r,s'\sim \mathcal{D}} \mathcal{T}^2(\theta)\\
    \mathcal{R}(\theta)&=\mathbb{E}_{s\sim \mathcal{D},a\sim\pi_{e}(a|s)}\hat{Q}_{k}(s,a,\theta) - \mathbb{E}_{s,a\sim \mathcal{D}}\hat{Q}_{k}(s,a,\theta)\\
    \mathcal{T}(\theta) &=\hat{Q}_{k}(s,a,\theta) - \big [ r + \gamma \mathbb{E}_{s'\sim \mathcal{D},a' \sim \pi_{e}(a'|s')}\hat{Q}_{k}(s',a',\theta) \big ]
\end{split}
\end{equation}
where $\mathcal{R}(\theta)$, called CQL regularizer, is used to penalize Q values of OOD data and $\mathcal{T}(\theta)$ is the standard Temporal-Difference (TD) error. $\mathcal{D}$ is the testing dataset collected from our online recommendation platform, as mentioned in Section \ref{subsec:Experiment dataset}. $\alpha$ is the penalty factor that regulates the punishment degree to the CQL regularizer.
To reduce the estimated error, we set $\alpha$ as $0.5 \times 10^{-3}$  to learn $\hat{Q}_{k}$ such that the expected value of a policy $\pi_{e}(a|s)$ under this Q-function lower-bounds its true value. We describe the implementation details of Conservative-OPEstimator in Appendix \ref{app:ope}.

In our experiments, we take the output value of Conservative-OPEstimator in Equation (\ref{eq:OPE Q value}) as the offline evaluation metric.
Specially, we use \textbf{Long-term user satisfaction per session $\hat{V}(\pi_e)$} to measure the average present value of long-term user satisfaction generated by a recommendation policy during a session.

\subsubsection{Online A$/$B testing.}
We deploy baselines to a large short video platform over a month of online A$/$B test and focus on long-term user satisfaction with the recommendation policies. 
We select two online metrics as follows to comprehensively test the model from two aspects of user stickiness and user activeness:

\noindent $\bullet$ \textbf{App dwell time (ADTime)} is the average APP usage time for all users within a day.

\noindent $\bullet$ \textbf{User positive-interaction rate (UPIRate)} is the percentage of video plays with positive user interactions during a day.

\subsection{Compared Methods}\label{subsec:Baselines}
\subsubsection{Baselines.} We compare our model with the common non-reinforcement learning algorithms and the advanced reinforcement learning algorithms. 

\noindent $\bullet$ \textbf{Bayesian Optimization (BO)} fits the prior distribution of the objective function based on Gaussian Process Regression and samples the optimal weight with upper confidence bound.

\noindent $\bullet$ \textbf{Evolutionary Strategy (ES)} uses a simple network whose inputs are user state and outputs are personalized fusion weights. With the network, ES turns to search the optimal network parameters through natural gradient.

\noindent $\bullet$ \textbf{Twin Delayed Deep Deterministic Policy Gradient (TD3) \cite{dankwa2019twin}} is an advanced off-policy algorithm, which also learns two target networks to reduce the overestimation bias. 

\noindent $\bullet$ \textbf{UWAC+TD3} connects TD3 with Uncertainty Weighted Actor-Critic(UWAC) \cite{wu2021uncertainty} based on the well-established Bayesian uncertainty estimation methods, so that it can identify OOD training samples and reduce their weights to train a conservative Q function.

\noindent $\bullet$ \textbf{CQL+SAC \cite{kumar2020conservative}} combines CQL with Soft Actor-Critic (SAC) \cite{haarnoja2018soft}. By regularizing the Q value of OOD action-state pairs, CQL algorithm learns a conservative, lower-bound Q function to reduce extrapolation error. SAC maximizes the cumulative rewards as well as the entropy of the policy to increase the stochastic of the policy and break away from the local optimum. 

\subsubsection{Variations of our model.} \label{sec:model variants}
We also compare our model with two types of variants designed to illustrate the effects of online exploration and reward function on model performance, respectively.


\noindent $\bullet$ \textbf{BatchRL-MTF-RE} is trained only with the data from random exploration. We use this variant to illustrate the power of our mixed multi-exploration.


\noindent $\bullet$ \textbf{BatchRL-MTF-Rtime} aims to improve user stickiness by setting a larger affinity weight on video play time in Equation (\ref{eq:reward}).

\noindent $\bullet$ \textbf{BatchRL-MTF-Rintegrity} aims to improve user stickiness by setting a larger affinity weight on video play integrity in Equation (\ref{eq:reward}).

\noindent $\bullet$ \textbf{BatchRL-MTF-Rinteraction} aims to improve user activeness by setting larger affinity weights on interaction behaviors in Equation (\ref{eq:reward}).

\begin{figure}[t]
    \centering
    \includegraphics[width=7.5cm,height=5cm]{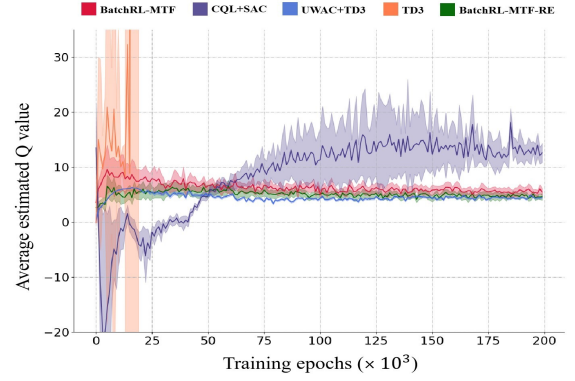}
    \captionsetup{font={small,bf},justification=raggedright}
    \caption{The estimated Q value curve of different policy.}
    \label{fig:Q value}
\end{figure}

\begin{table}[t]
\caption{Offline and Online Experimental results. The online experiment results represents the improvement of other methods on ADTime and UPIRate compared with Bayesian Optimization (BO). `*' means as the benchmark, BO results is 0. The best result is bold. `-' means that the result is empty. }
\vspace{-2mm}
\begin{center}
\begin{tabular}{c||c|cc}
\toprule
\multirow{2}{*}{\diagbox{Methods}{Metrics}} & offline & \multicolumn{2}{c}{online} \\ 
\cline{2-4}
& $\hat{V}(\pi_e)$ & ADTime & UPIRate \\ \midrule \midrule
BO & - & * & * \\
ES & - & +0.376\% & +0.402\% \\
TD3 & -648.162 & - & - \\
UWAC+TD3 & -297.053  & -0.513\% & \textbf{+16.061}\% \\
CQL+SAC & \textbf{5.194} & +2.322\% & +10.258\% \\
\midrule \midrule
BatchRL-MTF & 4.126 & +2.216\% & +9.118\% \\
BatchRL-MTF-RE & 3.023 & +0.862\% & -1.282\% \\
BatchRL-MTF-Rtime & - & +2.254\% & +8.877\% \\
BatchRL-MTF-Rintegrity & - & + 1.996\% & +9.464\% \\
BatchRL-MTF-Rinteraction & - & \textbf{+2.550\%} & +9.651\% \\
\bottomrule
\end{tabular}
\label{table2}
\end{center}
\end{table}


\subsection{Offline Evaluation} \label{sec:Offline Evaluation}
In this section, we conduct extensive offline experiments to verify the excellent performance of our Batch RL model with strong robustness, high returns and less extrapolation error, while demonstrating our online exploration that can discover high-value training samples to prevent policy from falling into local optimum. In addition, we test the sensitivity of parameters in our model and details of the analysis results are presented in Appendix \ref{app:par}.
\subsubsection{Effectiveness of Our Batch RL Framework} \label{sec:Offline RLmodel Evaluation}
\begin{figure*}[t!]
    \centering
    \includegraphics[width=16cm, height=4cm]{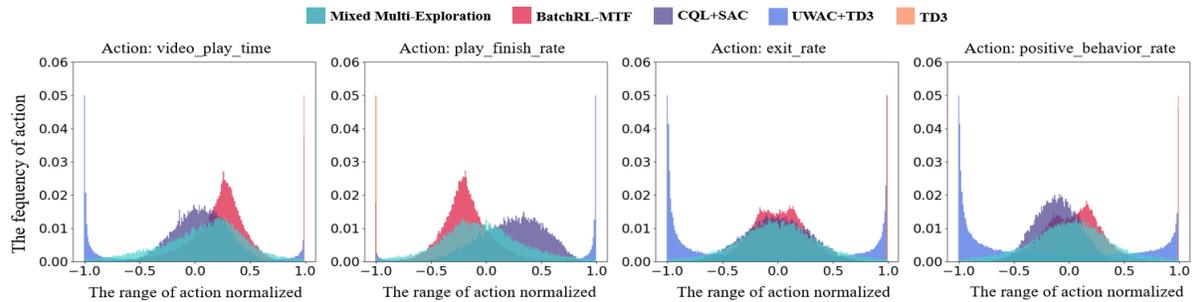}
    \captionsetup{font={small,bf},justification=raggedright}
    \caption{Action distribution of our model and other RL model of baselines. We select the most representative $4$-dimensional actions video play time, play finish rate, exit rate and positive behavior rate for comparison.}
    \label{fig:action 1}
\end{figure*}

As mentioned, the traditional off-policy algorithms exhibit large error in the estimation of Q value when we use the fixed dataset for offline learning. In Figure \ref{fig:Q value}, we show RL algorithms of the average estimated $Q$ value and its variance during training, and further show the policy action distribution in Figure \ref{fig:action 1} to observe extrapolation error of RL models. All models are trained by samples from mixed multi-exploration policy and all action values are truncated within a pre-defined interval and then normalized to $[-1,1]$. 

Extrapolation error is the overestimation bias to Q value of OOD data, and it is constantly accumulated by iterating the Bellman backup operator. We notice that TD3 tends to output extreme action values that are out of the distribution of batch data in Figure \ref{fig:action 1} because its Q function overestimates Q values of OOD actions. The estimated Q value of TD3 in Figure \ref{fig:Q value} exponentially increases in the early stage and can not converge to a stable and reasonable value. 
In order to alleviate extrapolation error of TD3, we also try to optimize TD3 with UWAC algorithm. In Figure \ref{fig:action 1}, different from TD3 which only generates OOD actions, UWAC+TD3 exploits Monte Carlo (MC) dropout to identify OOD actions and avoids learning from these actions. As a result, the model has lower probability to produce OOD actions. 
Unfortunately, UWAC+TD3 still have serious extrapolation error problem.

Both BatchRL-MTF and CQL+SAC perform well on reducing extrapolation error.
However, these two methods apply different strategies during training process.
CQL+SAC model chooses to penalize the Q values of unseen state and action pairs.
In complex RS with noisy user feedback, this soft constraint strategy may not get rid of all OOD actions.
Different from CQL+SAC, BCQ model introduces the action generative network that hard constraints the output actions around the seen ones and exploits online exploration to increase the diversity of actions.
In Figure \ref{fig:Q value}, we find that, although BCQ produces lower Q value, it is more stable than CQL+SAC and gives faster convergence during training, which is benefited from the direct constraints on actions.
In next section, we also show that BCQ achieves stable improvements on both ADTime and UPIRate on our online RS.

In addition, we take our Conservative-OPEstimator to evaluate all models and the result are shown in Table \ref{table2}. Conservative-OPEstimator gives a negative evaluation value $\hat{V}(\pi_e)$ to TD3 and UWAC + TD3 both with large extrapolation error, and thinks these models would hurt user experience. The above results also confirm our policy takes effect that Conservative-OPEstimator tries to punish the estimated Q values of OOD actions to 
avoid overestimation. Although CQL+SAC with a highest evaluation value, $\hat{V}(\pi_e)$, as an average metrics, can only represents the overall rewards. 
Compared to BCQ, which directly limits the distribution of actions, CQL+SAC has a more volatile output, which is fatal to large-scale personalized recommendation platforms that demand online stability. In a word, the overall performance of our model is optimal in considering of both stability and returns.

\subsubsection{Effectiveness of Online Exploration}

\begin{figure*}[t!]
    \centering
    \includegraphics[width=16cm, height=4cm]{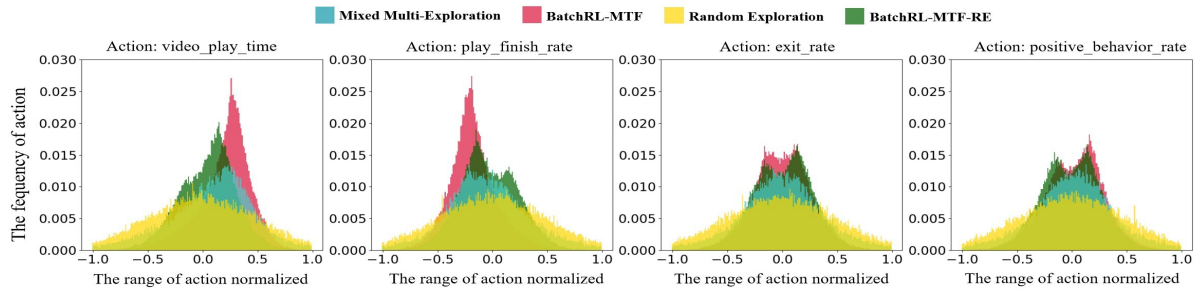}
    \captionsetup{font={small,bf},justification=raggedright}
    \caption{Action distribution of our model with different online exploration. We select the most representative $4$-dimensional actions video play time, play finish rate, exit rate and positive behavior rate for comparison.}
    \label{fig:action 2}
\end{figure*}

We also compare the output action distribution of our methods with different online exploration policies, i.e., BatchRL-MTF and BatchRL-MTF-RE, in Figure \ref{fig:action 2}.
Because Batch RL model based on behavior cloning constraints itself not to output OOD actions, the action distribution of both Batchr-MTF and Batchr-MTF-RE is more concentrated than that of their fixed batch data. However, the policy searches for the optimal action in the batch dataset, which limits the improvement of policy and causes the local optimal problem.

Therefore, our online exploration combines the extensive exploring of random exploration with the priori knowledge of action-noise exploration to construct high-quality dataset. 
It is obvious that the action distribution of BatchRL-MTF is more concentrated than that of behavior policy and BatchRL-MTF-RE. Compared with BatchRL-MTF-RE, BatchRL-MTF explores potential high-value state-action pairs based on the priori knowledge, which reduces those unnecessary exploration. 
The evaluation results of Conservative-OPEstimator in Table \ref{table2} also verify the effectiveness of our online exploration policy that focuses on searching for high-value samples while ensures the diversity of actions.
Later, we will further show the efficiency of our online exploration via an online comparison experiment.

\subsection{Online Evaluation}
In this section, we deploy all compared models and our BatchRL-MTF on an online RS, except TD3 (it doesn't work and could seriously hurt user experience). We not only compare our model with other baselines, but also discuss the weights $w_i$ in reward function on their effect of our model. 
\subsubsection{Comparison against Other Methods}
We use BO as the benchmark for comparison and show the improvements of ADTime and UPIRate for other methods and the experimental results are shown in Table \ref{table2}, where all improvements have statistical significance with $\text{p-value}$<$0.05$. 
ES slightly improves the performance by considering user preferences, but is still a performance gap to RL models that aim to optimize long-term user satisfaction. 
For the evaluation of RL models, the online evaluation results are completely consistent with the offline evaluation results given by our Conservative-OPEstimator. UWAC+TD3 performs poorly, dragging down ADTime by 0.513\%; the ADTime and UPIRate returns of CQL+SAC are both the highest in baselines, but not as stable as that of our model. The specific reasons for their performance have been analyzed in Section \ref{sec:Offline RLmodel Evaluation} and we will not repeat it.

Compared with BatchRL-MTF-RE, the mixed multi-exploration of our model significantly outperforms Gaussian Noise random exploration.
To be more specific, random exploration increases $0.862\%$ ADTime but decreases $1.282\%$ UPIRate.
The reason may lay in the large amount of noisy exploration on the whole action space that damaged the user experience.
On the contrary, our BatchRL-MTF explore around optimal actions and effectively learns the optimal policy from the high-value samples.

\subsubsection{Affinity Weight in Reward Function}
To explore the impact of different affinity weights $w_i$ in reward function, we deploy variants described in Section \ref{sec:model variants} on our RS and compare real user feedback.
As shown in Table \ref{table2}, compared with BatchRL-MTF, BatchRL-MTF-Rtime only gives a slight improvement on user stickiness, i.e., increasing ADTime by 0.037\%; while harms user activeness significantly, i.e., decreasing UPIRate by 0.241\%.
Similarly, BatchRL-MTF-Rintegrity also exhibits a seesaw phenomenon between user stickiness and activeness.
Clearly, the weights increasing of video play time or integrity may misleading the agent.
For example, the agent will recommend more long videos to increase video play time or recommend more short videos to improve video play integrity.
We notice that BatchRL-MTF-Rinteraction performs best among these variants, which improves ADTime and UPIRate by 0.333\% and 0.534\% respectively.
The probable reason may be that user interaction behaviors are sparse but strong signals which can guide the agent to learn user preferences and improve user satisfaction.

\section{Related Work}
Although there have been extensive studies on reinforcement learning based recommender systems \cite{dulac2015deep,zhao2018deep,wang2018supervised,afsar2021reinforcement}, we note that their problems are different from those studied in this paper.
Particularly, the output (action) of the above methods is usually recommendation item(s), while in this paper we aim to figure out the optimal weights for model fusion of MTL predictions.

To figure out the optimal fusion weights, early works try to solve this problem via parameter searching algorithms such as Grid Search, Genetic Algorithm \cite{mitchell1998introduction}, and Bayesian Optimization \cite{movckus1975bayesian}.
For example, \cite{rodriguez2012multiple,gu2020deep} use Grid Search to iterate over all combinations of candidate parameter sets and select the optimal weights through A/B Test.
Galuzzi et al. \cite{galuzzi2020hyperparameter} propose to use Bayesian Optimization to optimize the number of latent factors, the regularization parameter, and the learning rate.
The main drawback of these methods is that they always produce unified fusion weights across different users and thus can not model user preferences.
To search personalized fusion weights, Ribeiro et al. \cite{ribeiro2014multiobjective} propose using evolutionary algorithm to find Pareto-efficient hybrids.
However, all above methods focus on instant returns but ignore long-term user satisfaction.

Recently, to maximize long-term user satisfaction, a few works try to search the optimal weights via reinforcement learning.
Pei et al. \cite{pei2019value} propose a reinforcement learning based model to maximize platform expected profit.
To simplify the model, they use evolutionary strategy to solve the problem and thus the proposed method are still limited to optimize the profile of current recommendation.
Han et al. \cite{han2019optimizing} exploit off-policy reinforcement learning to find out the optimal weights between the predicted click-through rate and bid price of an advertiser.
Because numerous interactions with the immature agents will harm user experiences, they build an environment simulator to generate users' feedback for training their model offline.
However, the real recommendation environment is so complex that the simulator can't simulate it completely. In fact, the RL model based on the simulator will be difficult to adapt to the online environment and hurt user experience.

\section{Conclusion}

Multi-Task Fusion (MTF), which determines the final recommendation, is a crucial task in large RSs but has not received much attention by far in both academia and industry.
In this paper, we propose BatchRL-MTF framework for MTF recommendation task, which aims to optimize long-term user satisfaction.
We first formulate our recommendation task as an MDP and deliberately design the reward function involving multiple user behaviors that represent user stickiness and user activeness.
In order not to hurt online user experience, we exploit Batch RL model to optimize accumulated rewards and online exploration policy to discover potential valuable actions.
Finally, we propose a Conservative-OPEstimator to test our model offline, while conduct online experiments in real recommendation environment for comparison of different models. Experiments show that our model has the advantages of high returns, strong robustness and small extrapolation error. In addition, we also explore the effect of affinity weight in the reward function on our model, and find that when the weight of user activeness feedback is increased, our model will obtain higher returns. 
Furthermore, we successfully implement our model in a large-scale short video platform, improving 2.550\% app dwell time and 9.651\% user positive-interaction rate.

\bibliographystyle{ACM-Reference-Format}
\balance
\bibliography{reference}

\appendix
\newpage
\section{Implementation Details of Conservative Offline Policy Estimator}\label{app:ope}
In Algorithm \ref{alg:ope}, we elaborate the training and evaluation steps of our Conservative-OPEstimator. The Q network in Conservative-OPEstimator are MLP with ReLU activation function in hidden layers and are optimized based on Adam optimizer with the learning rate set to $0.1 \times 10^{-3}$. In addition, we set the training batch size, the testing batch size, the discount factor, the penalty coefficient and training epochs to $m=512 , n=5000 , \gamma=0.95$, $\alpha=0.5 \times 10^{-3}$ and $K=5000$, respectively.

\begin{algorithm}[t]
	\SetKwInOut{Input}{Input}
	\SetKwInOut{Output}{Output}
	\BlankLine
 	\Input{the transition dataset $\mathcal{D}=\{s_{i}, a_{i}, r_{i}, s'_{i}\}_{i=1}^{N}$, training batch size $m$, testing batch size $n$, discount factor $\gamma$, the penalty coefficient $\alpha$, the policy $\pi_{e}$ to be evaluated.}
 	\BlankLine
	
	Initialize $Q_{0}(\cdot,\theta)$ randomly;
	
	
    	\ForEach{$0 \le k \le K$}{
    	    Sample training batch of $m$ transitions $\{s_i,a_i,r_i,s'_i\}_{i=1}^m$ from $\mathcal{D}$ randomly\;
    	    Compute current $\hat{Q}_{i}^e=Q_{k}(s_i,\pi_e(s_i))$ and $\hat{Q}_{i}=Q_{k}(s_i,a_i)$, target $\hat{y}_{i}=r_{i}+ \gamma Q_{k}(s'_i,\pi_e(s'_i)) \quad  \forall i$\;
            Construct training batch set:
            \begin{equation}\label{eq:OPE Q}
            \begin{aligned}
                \widetilde{\mathcal{D}}_k=\{(s_i,a_i), \hat{Q}_{i}^e, \hat{Q}_{i}, \hat{y}_{i}\}_{i=1}^m; \nonumber
            \end{aligned}
            \end{equation}
            \\
            Fit Q-function based on regression:
            \begin{equation}
            \begin{split}
                Q_{k+1} \gets& \mathop{argmax}_{\theta} \alpha \cdot \left( \dfrac{1}{m}\sum_{i=1}^{m} \hat{Q}_{i}^e - \dfrac{1}{m}\sum_{i=1}^{m} \hat{Q}_{i} \right ) \\
                & +{\frac{1}{2}} \cdot \dfrac{1}{m}\sum_{i=1}^{m} \left( {Q}_{k}(s_i,a_i)-\hat{y}_i \right )^2; \nonumber
            \end{split}
            \end{equation}
            \\
        }
    Obtain fitted Q-function $\hat{Q}(\cdot,\theta)=\lim\limits_{k \rightarrow K}\hat{Q}_{k}$\;
    Sample testing batch of $n$ initial states $\{s_i^0\}_{i=1}^n$ from $\mathcal{D}$ randomly\;
    Evaluate the offline policy $\pi_e$:
    \begin{equation}\label{eq:OPE}
        \hat{V}(\pi_e) = \dfrac{1}{n}\sum_{i=1}^{n}\sum_{a \sim \pi_{e}(a|s) }\pi_{e}(a|s^0_i)\hat{Q}(s^0_i,a). \nonumber
    \end{equation}

 	\Output{the offline evaluation result $\hat{V}(\pi_e)$.}

	\caption{Offline Evaluation Steps of Conservative Offline Policy Estimator $V(\pi_{e})$.\label{alg:ope}}
\end{algorithm}

\section{Sensitivity Analysis of Parameters}\label{app:par}
In this section, we will observe the influence on our model performance of two important model parameters, the perturbation bound $\rho$ and the learning rate $\eta$ of the critic network, to figure out their optimal values. We evaluate Batch RL model class with the different value of $\rho$ or $\eta$, and visualize the offline evaluation results in Figure \ref{fig:phi}.
\subsection{Sensitivity of the Perturbation Bound $\rho$}
As the behavior cloning-based method, Our Batch RL model reduces extrapolation error by constraining itself not to output OOD actions. It is essentially a fine-tuned policy based on training data generated by behavior policy. Once the batch data set constructed by the behavior policy is not the optimal set, our model will fall into the dilemma of local optimum. In order to explore potential high-value state-action pairs but avoid the action with noise to hurt user experience, we exploit the action perturbation network $P_{\omega}(s,a,\rho)$ to perturb actions cloned by VAE model, as discussed in Section \ref{sec:actor}. The perturbation bound $\rho$ is a critical parameter that controls the range of exploration and influences the final output action. 

As shown in Figure \ref{fig:phi} (a), when $\rho=0.15$, our model has the best performance and can achieve stable and high returns. With the decrease of $\rho$, the exploration for the action is conservative and narrow, which is conducive to obtain the model with stable returns but limits the model to seeking high returns; Instead, the exploration is more random and extensive, which is conducive to enrich the action candidate set but increases extrapolation error and the action with noise of the model. The above phenomenon also shows that $\rho$ only adjusts the constraint range of the model on the output action, and the Q network of our model still overestimates the Q value of OOD actions. A large $\rho$ means that there is a small constraint on the action. Therefore, our model may output OOD actions, resulting in extrapolation error. 

\subsection{Sensitivity of the Learning Rate $\eta$ of the Critic Network.}
The critic network of our Batch RL model, which can evaluate the value of state-action pairs, provides the actor network with the optimization direction and the basis for decision making. Therefore, it directly affects our model performance whether the critic network learns adequately and effectively. 

\begin{figure*}[t!]
    \centering
    \includegraphics[width=16cm]{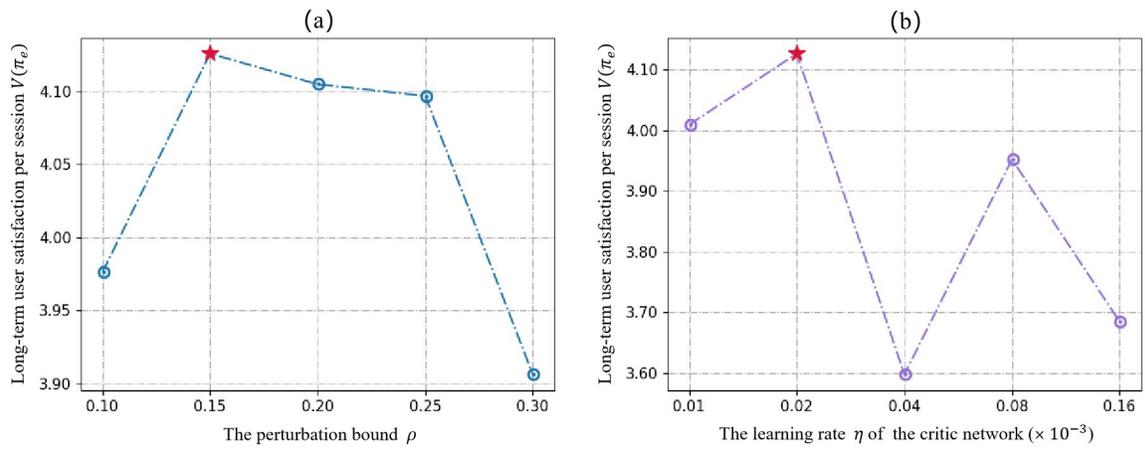}
    \captionsetup{font={small,bf},justification=raggedright}
    \caption{Parameter sensitivity of the perturbation bound $\rho$ and the learning rate $\eta$ of the critic network. The best result is marked by a red star.}
    \label{fig:phi}
\end{figure*}

In Figure \ref{fig:phi} (b), we research how the learning rate $\eta$ of the critic network influences the long-term rewards of our model. With the increase of learning rate $\eta$, the rewards of model fluctuates constantly. We note that, when the learning rate $\eta$ is around $0.2 \times 10^{-3}$, the fluctuation flattens out and our model gains the highest returns. This shows that it is so appropriate to set $\eta =0.2 \times 10^{-3}$ that the critic network does not fall into the local optimal because of a small $\eta$, nor miss the real optimal solution because of a large $\eta$. Meanwhile, the result ensures the critic network trained adequately that can accurately evaluate the value of state-action pairs.



\end{document}